\documentclass[useAMS,usenatbib]{mn2e} 

\usepackage{graphicx} \usepackage{txfonts} \usepackage{longtable}
\usepackage{lastpage}

\title[NGC~6302 Expansion proper motions]{The Expansion Proper Motions of the
Planetary Nebula NGC~6302 from HST imaging}

\author[C. Szyszka, A. A. Zijlstra and J. R. Walsh] { C.
Szyszka$^{1}$\thanks{E-mail:cszyszka@gmail.com (CS)}; A. A. Zijlstra$^{1}$ and
J. R. Walsh$^{2}$ \\ $^{1}$School of Physics \& Astronomy, University of
Manchester, Manchester M13 9PL, UK\\ $^{2}$European Southern Observatory,
Karl-Schwarzschild Strasse 2, D85748 Garching, Germany}

\begin{document}

\date{Accepted 2011 May 16.  Received 2011 May 12; in original form 2011 February 24}


\maketitle

\label{firstpage}

\begin{abstract} Planetary nebulae expand on time scales of $10^3$--$10^4$\,yr.
For nearby objects, their expansion can be detected within years to decades. The
pattern of expansion probes the internal velocity field and provides clues to
the nebula ejection mechanism.  In the case of non-symmetric nebulae, and
bipolar nebulae in particular, it can also provide information on the
development of the morphology.  We have measured the expansion proper motions in
NGC~6302 from two epochs of HST imaging, separated by 9.43 years. This is used
to determine the expansion age and the structure of the velocity field.  We use
HST images in the [N~II] 6583\AA\ filter from HST WF/PC2 and WFC3. The proper
motions were obtained for a set of 200 individual tiles within 90$"$ of the
central star. The velocity field shows a characteristic linear increase of
velocity with radial distance (a so-called Hubble flow). It agrees well with a
previous determination by Meaburn et al. (2008), made in a lobe further from the
star, which was based on a much longer time span, but ground-based imaging. The
pattern of proper motion vectors is mostly radial and the origin is close to the
position of the central star directly detected by Szyszka et al. (2009).  The
results show that the lobes of NGC~6302 were ejected during a brief event
$2250\pm35$yr ago. In the inner regions there is evidence for a subsequent
acceleration of the gas by an additional 9.2 km/s, possibly related to the onset
of ionization.  The dense and massive molecular torus was ejected over 5000yr,
ending about 2900yr ago. The lobes were ejected after a short interlude (the
'jet lag') of $\sim 600$\,yr during a brief event. The torus and lobes originate
from separate mass-loss events with different physical processes.  The delay
between the cessation of equatorial mass loss and the ejection of the lobes
provides an important constraint for explaining the final mass-loss stages of
the progenitor stellar system. \end{abstract}


\section{Introduction} Planetary nebulae (PNe) are among the fastest evolving
objects in the Universe.  They form when a solar-like star with mass in the
range 1--7 M$_\odot$ reaches the asymptotic giant branch (AGB).  On the AGB the
star develops a degenerate C/O core and a convective envelope which is ejected
during a phase of extreme mass loss \citep[e.g.][]{Habing:1996}. The ejected
envelope expands away from the star with velocities of a few tens of
km\,s$^{-1}$.  The remaining core quickly transitions from a surface temperature
of $\sim 3 \times 10^3$\,K to $\sim 10^5$\,K, within $\sim 10^4$\,yr, before
nuclear burning ceases and the star enters the white dwarf cooling track. The
envelope is ionised by the hot core and forms the visible PN.

Since the shell ejection is short-lived, expansion of the planetary nebula can
be observed within decades for nearby PNe. This expansion in the plane of the
sky (expansion proper motion) has been measured for several objects, mostly
using Very Large Array (VLA) and/or Hubble Space Telescope (HST) observations. 
By equating the nebula expansion in the plane of the sky with the measured
line-of-sight expansion velocity, expansion proper motions have been used as a
tool for measuring distances \citep{Masson:1986, Hajian:1993, Meaburn:1997,
Palen:2002, Meaburn:2008, Guzman:2009} and thus stellar masses
\citep{Zijlstra:2008}.

The expansion of the planetary nebulae is assumed to be spherically symmetric
\citep[e.g.][]{Guzman:2009} or described by a simple velocity law
\citep{Zijlstra:2008}. But PNe show complex morphologies, including tori,
bipolar outflows, highly collimated jet-like structures, and knots
\citep{Balick:2002}. It is unlikely that all these structures expand in
spherical unison. In fact, velocity fields and expansion proper motions in some
cases show evidence of Hubble-like (viz. $ V \propto r$) outflows
\citep[e.g.][]{Meaburn:2008, Corradi:2004}. High accuracy expansion measurements
can reveal the internal dynamics of the complex structures, and thus help to
constrain their origin and evolution.

A detailed study of the expansion of the the planetary nebula NGC\,6302 (PN
G349.5+01.0) is presented here based on the highest spatial resolution imagery
available. NGC~6302 is among the most strongly bipolar (or even multi-polar) of
the Galactic planetary nebula, and has been imaged twice with the HST.

\citet{Meaburn:2008} measured the proper motion of 15 features (knots) in the
north-western lobe, from two observations separated by a period of 50.88 years. 
From the expansion parallax, a distance of 1.17\,kpc and an age of 2200 years
were derived. The earlier epoch observation was based on archival photographic
plates \citep{Evans:1959}. The accuracy is limited by (apart from the time
baseline), the image quality of the plate, and the seeing of the ground-based
observations. To significantly improve on these data, space-based observations
are required.

Two sets of HST observations taken $\sim$\,9.4 years apart are used in this
paper to study the differential expansion of the lobes in the inner regions of
the nebula.  The HST images give a spatial resolution of 0.1$''$ per pixel,
allowing very compact knots and structures to be resolved, and to thus detect
small proper motions.  Section 2 presents the observations. The reduction and
analysis are described in Section 3. The discussion in section 4 concentrates on
the implication of the pattern of the nebula expansion proper motions.
Conclusions are presented in Section 5.

\section{Observations} The planetary nebula NGC\,6302 has been imaged with the
Hubble Space Telescope (HST) at two epochs. The first data set was taken on 2000
February 21 (MJD = 51595.93171041) with Wide Field Planetary Camera 2 (WF/PC2).
Two narrow-band filters F658N and F656N were used for this observation.  The
second set of imaging observations were made on 2009 July 27 (MJD =
55039.80441102) with Wide Field Camera 3 (WFC3) ultra-violet and visible channel
(UVIS) as part of the Servicing Mission 4 Early Release Observations. Among the
six narrow-band filters used, F658N, F656N are in common with the WF/PC2 data.
Table \ref{tab:observations} presents brief details of both sets of
observations.

\begin{figure} \centering \includegraphics[width=0.45
\textwidth]{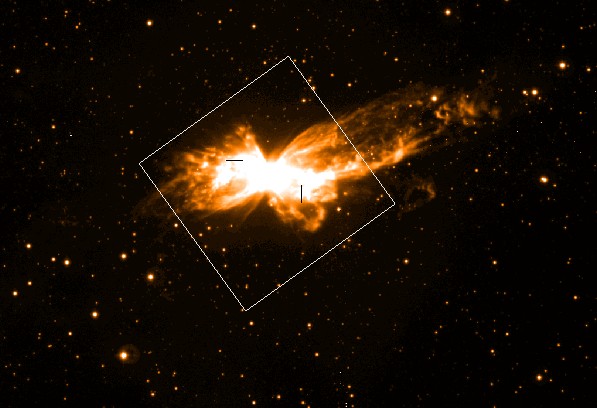} \caption{The approximate area covered by the
WFC3 images ($162\times162$ arcsec) is shown by the box, overlaid on a
wide-field H$\alpha$ image of NGC~6302, taken with the Wide Field Imager (WFI)
instrument at the ESO 2.2\,m telescope.  North is up and East is left. The
outermost lobes extend over 7 arcmin on the sky.} \label{fig:wfi} \end{figure}

The existence of two epochs of observation, separated by 3443.8727 days (9.4353
years) presents the opportunity to directly measure proper motions (PM) of the
expanding nebula. The WFC3 field of view is $162 \times 162$ arcsec. This covers
the bright, central region of the nebula. The approximate area covered is shown
in Fig \ref{fig:wfi}, overlaid on an ground-based H$\alpha$ image covering the
full nebula. The older WF/PC2 observations \citep{Matsuura:2005}, taken with a
smaller field of view, focussed on the eastern lobe.  The WFC3 data cover most
of the nebula, but the expansion could obviously only be studied in the region
of overlap.  \citet{Meaburn:2008} used the opposite, western lobe, and used data
at larger distances from the central star than studied here.

\begin{figure} \centering \includegraphics[width=0.45
\textwidth]{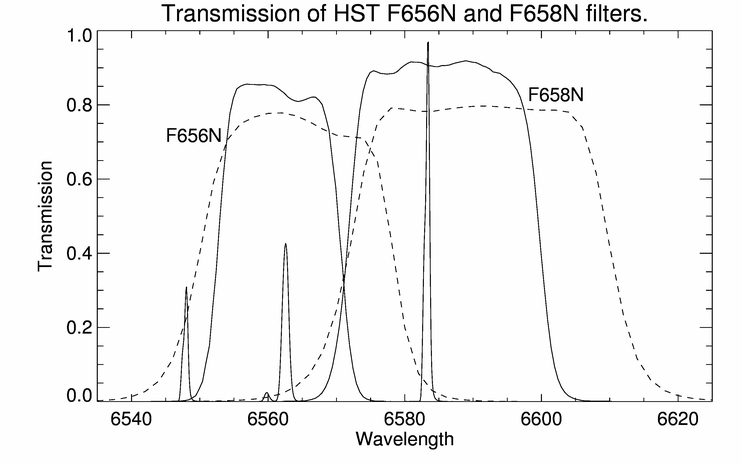} \caption{The transmission curves for filters used in
HST WF\/PC2 and WFC3 UVIS cameras. The dashed and the solid lines correspond to
WF\/PC2 and WFC3 filters respectively. A reference spectrum of NGC~6302 is also
presented, showing the two [N {\sc ii}] lines and the H$\alpha$ line (flanked by
a weak HeII line).} \label{fig:filters} \end{figure}

\begin{table} \caption{Summary of HST imaging observations of NGC~6302 in 2000
and 2009.} \label{tab:observations} \centering \begin{tabular}{c c c c}
\hline\hline Filter & Date & Integration & Dataset \\ \hline F656N & 2000-02-21
& 610s & U5HH0602B \\ F658N & 2000-02-21 & 470s & U5HH0602R \\ F656N &
2009-07-27 & 2100s & IACO01010 \\ F658N & 2009-07-27 & 2220s & IACO01020 \\
\hline \end{tabular} \end{table}

\subsection{Differences in filters transmissions}
\begin{table} \caption{HST narrowband filter transmissions at the wavelengths of
the main emission lines } \label{tab:filters} \centering \begin{tabular}{l c c c
c} \hline\hline Instrument& Filter & [N {\sc ii}] 6549\AA\ & H${\alpha}$6565\AA\
& [N {\sc ii}] 6585\AA \\ \hline WF/PC2 &F656N & 0.22$^{\mathrm{a}}$ & 0.77 &
0.03 \\ WF/PC2 &F658N & -- & 0.03 & 0.78 \\ WFC3 &F656N & 0.01$^{\mathrm{b}}$ &
0.81 & -- \\ WFC3&F658N & -- & -- & 0.91 \\ \hline \end{tabular}
\begin{list}{}{} \item[$^{\mathrm{a}}$] The filter curve is steep at this
wavelength. Based on the largest observed velocity shifts on the nebula, the
transmission can vary within the range 0.1--0.4. \item[$^{\mathrm{b}}$] Similar
as for $^a$, the transmission ranges from 0.001 to 0.03. \end{list} \end{table}
During both epochs, NGC\,6302 was observed with both the F658N and F656N
filters.  Although the filter names stayed the same, the characteristics have
changed.  The WFC3 filters are narrower, sharper, and more efficient than the
WF/PC2 filters of the same name, as demonstrated by Figure \ref{fig:filters}. 
The F656N image is usually called the H$\alpha$ image while the F658N frame is
attributed to [N {\sc ii}] emission at wavelength 6583\AA.  For the expansion
measurements the [N {\sc ii}] image (F658N) proved to be most suitable and is
the one used in this work.

For reference a scaled spectrum of NGC\,6302 is shown in Figure
\ref{fig:filters} where three brightest emission lines are clearly visible
(H$\alpha$ 6563\,\AA, [N\,{\sc ii}] 6548 and 6583\,\AA).  The filter
transmissions at these specific wavelengths are listed in Table
\ref{tab:filters}.  These filter responses, together with the observed
integrated line flux ratios [N\,{\sc ii}](6549\,\AA):H$\alpha$:[N\,{\sc
ii}](6549\,\AA) = 1:1.6:3.0 \citep{Tsamis:2003} reveals that about 25\% of the
total photon count in the WF/PC2 F656N image comes from the [N{\sc ii}] doublet,
while the same ion contributes only about 2\% of the total flux for the WFC3
F656N filter.  This makes it more difficult to directly compare these two
H$\alpha$ images.  The nebular line wavelengths also depend on the position
within the nebula, due to the velocity field \citep{Meaburn:1980b,Meaburn:2005}.
This introduces a further transmission uncertainty for the [N {\sc ii}]
6548\,\AA line which is situated at the edge of the F656N filter.

The difference in the [N {\sc ii}] contribution causes a notable difference in
the two H$\alpha$ images. The WFC3 image, which isolates H$\alpha$ much better,
shows a smooth emission structure. The WFPC2 image shows a mixture of this
smooth component and a knotty component. This knotty component dominates the
F658N images and is attributed to the [N {\sc ii}] emission. Because of this
strong difference between the two H$\alpha$ images, and the fact that proper
motions measurements are aided by the presence of compact structures, we only
make use of the F658N images here.
\begin{figure} \centering \includegraphics[width=0.45
\textwidth]{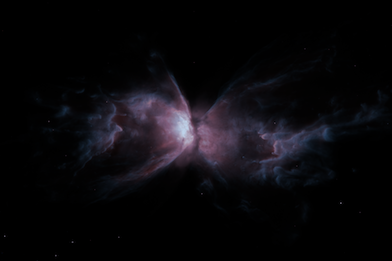} \caption{The WFC3 image, oriented such that
North is up and East is left.  The F658N image is blue and the F656N image is
red. } \label{fig:merge} \end{figure}

A two-colour figure obtained from the WFC3 F656N and F658N images is shown in
Fig. \ref{fig:merge}. It shows the complex structure, with the North-South
obscuring torus \citep{Matsuura:2005} and the well-defined edges in the lobes.

\section{Reduction and analysis} The analysis started with the pipeline-reduced
data products, provided by the Hubble Archive hosted by the Space Telescope
European Coordinating Facility (ST-ECF). For WF/PC2, data products from version
2.5.1 of the STSDAS calibration pipeline (CALWFP) were used while for WFC3 UVIS,
the data were calibrated using version 2.0 of the calibration pipeline (CALWF3).
The pipelined images were 'drizzled' (i.e. created by combining images with
sub-pixel shifts) using the task 'multidrizzle' in STSDAS. This also corrects
for the geometric distortion of the respective cameras.

Further reductions to improve the astrometric calibration, and the geometric
distortions were applied and are detailed below.

\subsection{Astrometry Calibration} The absolute astrometry as returned by the
pipeline reduction is based on guide stars from the GSC2, which have a typical
positional uncertainty of $0.2$--$0.28''$ depending on magnitude
\citep{Lasker:2008}.  Depending on the sets of guide stars used, the astrometry
for both sets of observations can differ. In addition the WFC3 data had not been
fully calibrated at the time the observations were reduced, so larger errors may
be present in the pipeline calibration.

To improve the relative astrometry of the two sets of images, we identified
common field stars present in both epochs. The stars were identified from the
Two Micron All Sky Survey (2MASS) catalogue \citep{Skrutskie:2006}, being the
most complete stellar catalogue at the relevant magnitude levels. Catalogue
stars were excluded where two sources were visible in the HST images within a
circle of 1$''$ radius from the 2MASS position (which would be merged at the
2MASS resolution).  Stars were also excluded if they were farther than 1$''$
from the 2MASS position, taken as evidence for large proper motion.

For the WF/PC2 frame, the astrometric solution was tested using the astrometry
tool in Gaia/Starlink with the 2MASS catalogue in the field of view (rejecting
galaxies). The solution showed an individual scatter in the stellar positions of
around 0.3$''$ and a mean shift of 0.1$''$ ($\approx$~1 pixel). The procedure
was repeated with the WFC3 image. This latter step revealed a distinct $\sim 3
''$ shift (about 75 WFC3 pixels), attributed to the pipeline reduction of these
early in-flight data. This shift was corrected within Gaia by a translation in
RA and Dec, assuming no rotation and a constant pixel scale.

At this point, the WFC3 and WFC2 images have the same astrometry but different
pixel scale and orientation.  We rotated the images to a N-E orientation (for
the WF/PC2 image this was already done by the pipeline), and re-binned the
higher resolution WFC3 data (0.04 arcsec/pixel) to the resolution of WF/PC2
chips (i.e. 0.1 arcsec/pixel), bringing both frames onto the same spatial grids.
 This was done utilising the MONTAGE software tool.

\begin{figure} \centering \includegraphics[width=0.45
\textwidth]{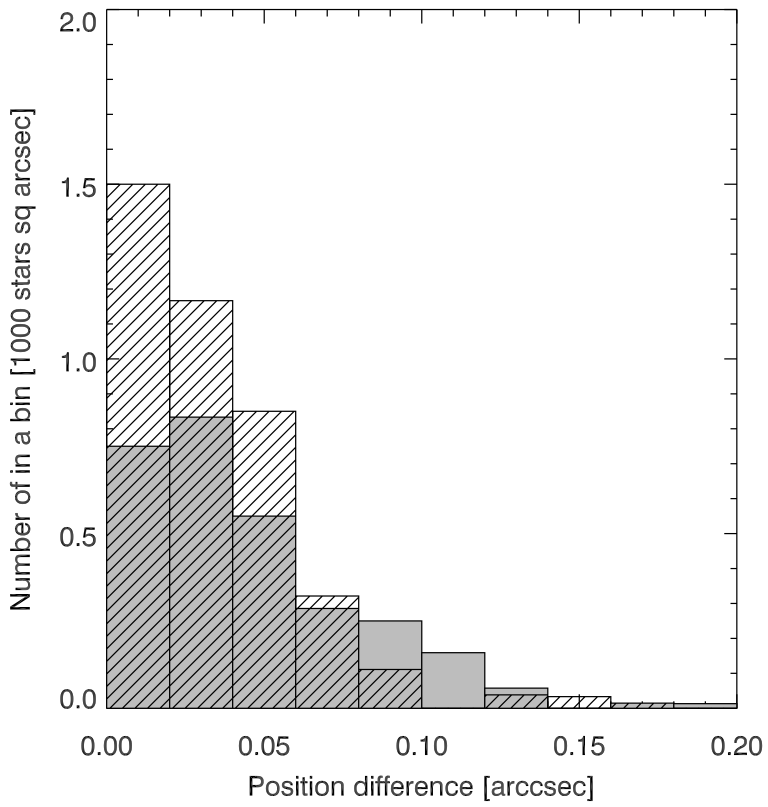} \caption{Histogram of the measured stellar
position differences between WF/PC2 and WFC3 UVIS images. The hatched region
shows the histogram of the position differences before correction for shear (see
Fig. \ref{fig:distortion}), and the shaded region after correction for shear. A
difference of 0.1$''$ is equal to 1 pixel.} \label{fig:histogram} \end{figure}

In a third step, for the 2MASS stars in common between both epochs, the centroid
positions were measured in GAIA/STARLINK in each frame to test for shifts. 
Sources which presented a shift between the two epochs larger than 1.4pix
(0.14") were excluded, suspected to be stars with high proper motion (this step
removed five stars).  This left 54 stars in common between both frames. The
histogram of the shifts between these 54 stars is shown in Fig. 
\ref{fig:histogram} (the shaded area). This distribution is rather wide,
suggesting a possible residual astrometric error.

\begin{figure} \centering \includegraphics[width=0.45
\textwidth]{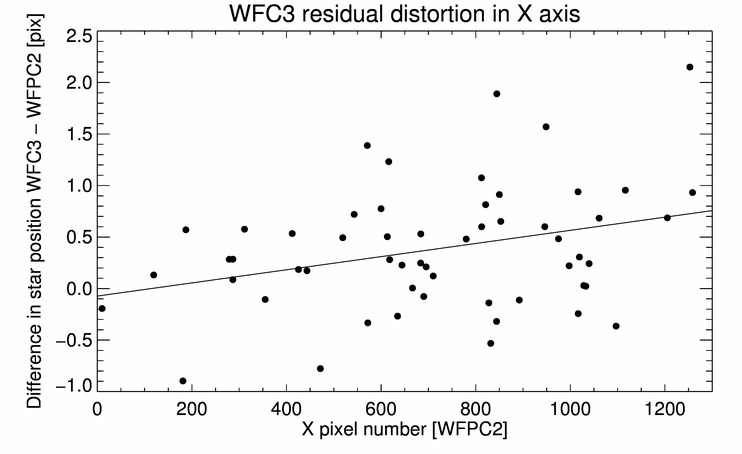} \includegraphics[width=0.45
\textwidth]{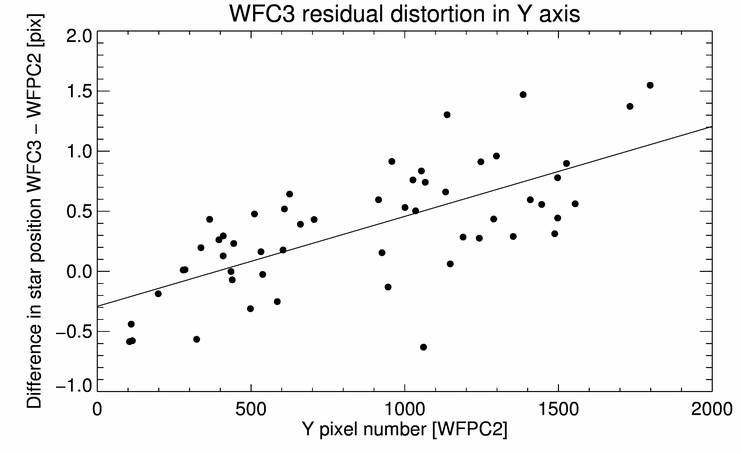} \caption{Distortion of the WFC3 UVIS
relative to the WF/PC2 as revealed by comparison of the positions of stars in
common on both images.} \label{fig:distortion} \end{figure}

\begin{figure} \includegraphics[width=0.45 \textwidth]{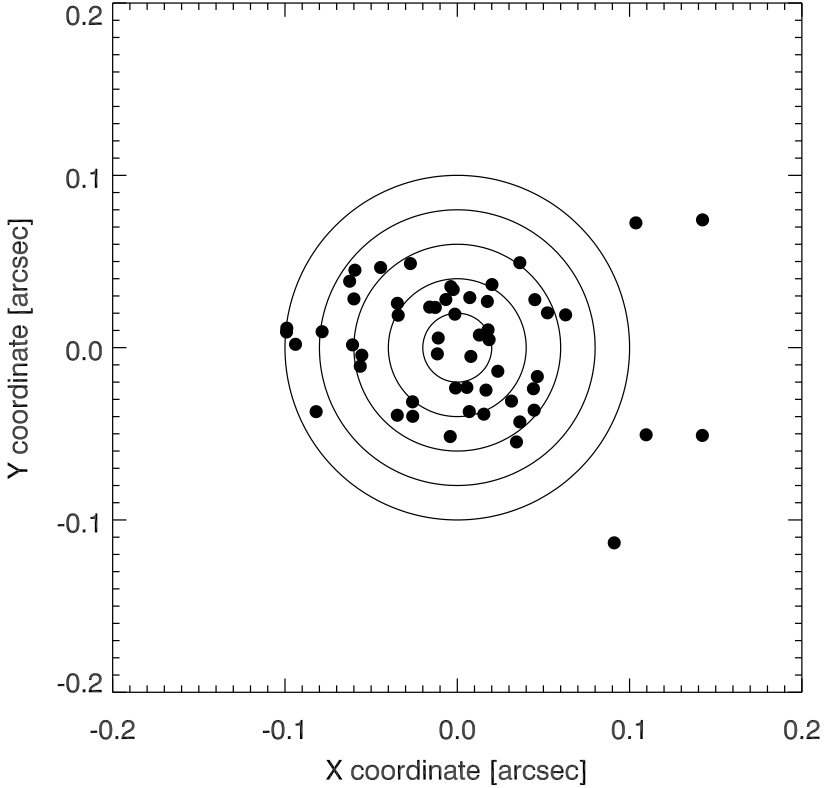}
\caption{Measured difference of the positions of individual stars between the
two epochs, after correction for shear. The concentric circles correspond to the
inner five bins of the histogram of Figure \ref{fig:histogram}.}
\label{fig:posdiff} \end{figure}

Comparing the difference in the positions of the stars on the WFC3 and WF/PC2
images as a function of $x$ and $y$ pixel number, shows that there is indeed a
residual distortion in one (or both) of the images, with an amplitude of about
0.1$''$ from edge to edge.  The shear between both sets of images is clearly
visible in Fig. \ref{fig:distortion}. We fitted the difference in star position
versus pixel number, separately for $x$ and $y$, with a linear functions of
pixel number ($x'-x=a_1x + b_1$ and $y'-y=a_2y + b_2$) where the primed
coordinates are the corrected ones. The best fit yielded
($a_1,b_1$)=($0.0006378,-0.0727$) and ($a_2,b_2$)=($0.0007481,-0.2905$). The
reliability of this transformation is limited by the intrinsic proper motion of
the stars which is not known. The constants $a_1$, $a_2$ indicate a small
difference in pixel scale, which needs to be corrected: the fact that both axes
yield the same factor within the uncertainties gives us confidence that this is
valid. The $b_1$, $b_2$ show a misalignment between the frames, largest in $y$. 
We applied the $a$ and $b$ coefficients to the WFC3 image.

This correction improved the astrometric calibration, as demonstrated by the
hatched histogram in Fig. \ref{fig:histogram}.  The mean scatter in stellar
positions is now better than 0.05$''$ (Fig. \ref{fig:posdiff}). This is probably
dominated by the proper motions of the individual stars. The motion corresponds
to about 25 km s$^{-1}$ at a distance of 1 kpc which is of the order of the
expected stellar velocity dispersion in the Galactic plane. The accuracy of the
relative alignment between the two images, as measured from the error on the
mean, is approximately 0.01\,arcsec.

\subsection{Proper motion measurements}

\subsubsection{Expected shifts} The expected level of proper motions can be
estimated from the existing measurements of \cite{Meaburn:2008}. They detected
proper motions in the northwestern lobe (see Fig \ref{fig:wfi}), between 1
arcmin and 3 arcmin from the centre of the nebula.  Their innermost measurement
derives from a feature at 1.06\,arcmin from the centre of the nebula, and
revealed a proper motion (PM) of 29.5\,mas/yr. At large distances, they find a
linear increase of PM with distance.

The most distant features measured in this work extend up to about 1$'$ from the
centre of the nebula, matching well with the innermost measurement of
\cite{Meaburn:2008} (albeit at opposite side of the nebula). The two epochs of
HST imaging are separated by $\approx$10 years. Thus, we expect shifts up to
0.3$''$ in the outer area of the HST images, down to 0.1$''$ at 20$''$ from the
centre. The latter corresponds to 1 pixel, and is ten times larger than the
alignment errors derived in the previous section.

\subsubsection{Method} 
\begin{figure} \centering \includegraphics[width=0.45
\textwidth]{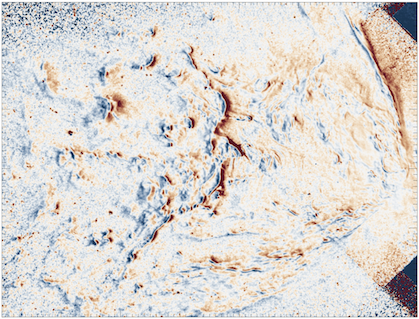} \caption{The difference image between the WF/PC2
and WFC3 F658N images, oriented such that North is up and East is left. The CCD
layout of the WF/PC2 is visible, with the smaller PC image on the right.  Blue
shows excess emission in WFC3, and red in WF/PC2. The shifts increase towards
the left, further from the centre.  } \label{fig:diff_image} \end{figure}

Direct evidence for proper motion of the structures in
the nebula can be seen from difference imaging. This is shown in Fig.
\ref{fig:diff_image}. The pattern shows clear motion of sharp-edged features, in
most cases directed away from the centre of the nebula. In the display, we used
the arctan of the differential intensity, to flatten the image. In reality, the
amplitudes in the right part, close to the center, are much higher than on the
left. The strong effect seen on the northern edge of the obscuring torus (top
right) is not caused by a very large motion, but represents a braided structure
with one chain brightening and the other fading. The intensity changes may be
partly caused by the H$\alpha$ emission in this region, which affects the two
filters differently. The results below indicate that the gas flow in this region
is along the linear feature.

The proper motion cannot be directly measured from the difference image.  In
regions where there is a field star close to a nebular feature, the motion can
be measured with respect to the reference star. In areas without such a
reference point a more detailed comparison is required.  The accuracy of the
proper motion is limited by the fact that the nebular features are not point
sources, but are extended. This introduces uncertainties which are much larger
than the systemic error derived from stellar (point) sources.

The best measurement for the proper motion is obtained by shifting the two
images and calculating the $\chi^2$ of the difference image. This technique
works for extended emission features, as long as the region studied is
sufficiently small that a linear shift provides a good approximation, whilst
there is sufficient structure in the emission features within this region. We
implemented this technique by considering the nebula as an array of adjoining
square tiles. To ensure sufficient structure within the selected region, we
chose to cut the nebula into a 23\,$\times$\,19 grid of fairly large square
tiles, 41\,$\times$\,41 pixels in size (4.1$''\times$4.1$''$) (437 tiles in
total.)

To determine the shift for each tile, we employed the MPFIT IDL package
\citep{Markwardt:2009}. This finds the best matching model iteratively by means
of $\chi^2$ minimization, with as parameters the shift in $x$ and $y$ (proper
motion of the nebular features) and the scaling factor for the intensity. The
scaling factor for the intensity accounts for the sensitivity difference between
WF/PC2 and WFC3. For each step (each shift) in this procedure, the WFC3 image
was interpolated to the pixel frame WFPC2 image, before the $\chi^2$ of the
difference image was calculated. This interpolation required non-integer shifts,
and was done using the IDL function INTERPOLATE.

The interpolation of the WFC3 image was always done using the entire WFC3 image.
This avoids edge effects on the extracted squares. The difference image was
obtained for the analysed tile only.

\subsubsection{Uncertainties} To assess the intrinsic errors of this method, we
artificially introduced a shift to the WF/PC2 image, and next used the same
method on the original and shifted WF/PC2 images. The aim was to reproduce the
shift. This measures both the intrinsic uncertainty in the calculation, and the
reliability of each tile.

We used shifts of ($-$2.0, $-$1.0, $-$0.8, $-$0.3) pixels in both directions,
extracted the full set of tiles, and re-ran the analysis for each tile. For
integer pixel shifts, the introduced shift was reproduced with accuracy close to
the computational precision for single-precision numbers, with standard
deviation of $10^{-5}$ pixels. (The standard deviation excludes tiles where the
calculated shift differed from the applied shift by more than 10\%.)

Not all tiles produce consistent results. Large errors are found for tiles which
lack nebular emission structure necessary for an accurate determination of the
shift.  We exclude tiles where the calculated shift was in error by more than
50\%\ for all of the trial shifts. This removed all tiles without detected
emission in the WF/PC2 image (the WFC3 image is deeper and shows emission in
some tiles which are empty in WF/PC2). It also removed a few tiles in regions
with smooth emission. It left 246 tiles.

Subsequently, we ran the full analysis on the WFC3 versus WF/PC2 image,
extracting the determined shifts for the tiles identified in the previous step.
A number of tiles were affected by field stars, which may move differently from
the nebula. These tiles (a total of 46) were manually removed. We also removed
tiles with very large shifts or where the shift measurements in one of the two
axes had failed. The final sample of 'good' tiles for proper motion measurements
is 200.

\begin{figure} \centering \includegraphics[width=0.45
\textwidth]{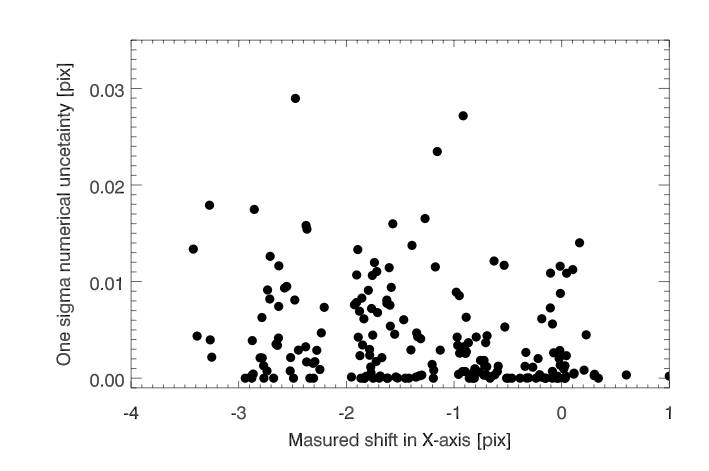} \includegraphics[width=0.45
\textwidth]{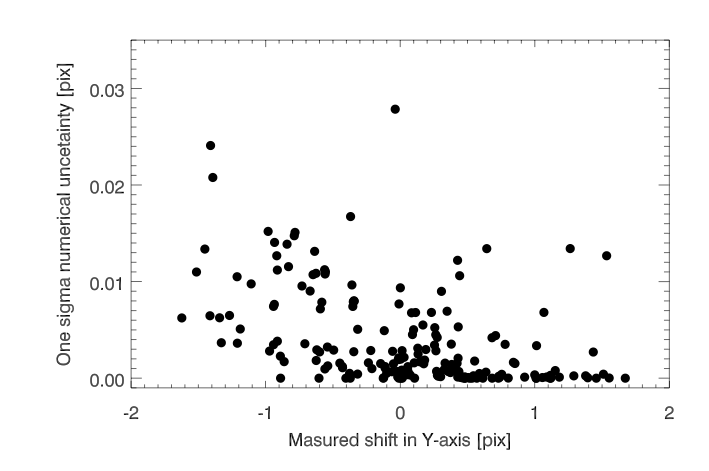} \caption{Formal numerical uncertainties,
as derived by MPFIT procedure, plotted versus measured shift for x (top) and y
(bottom) axes. } \label{fig:rawdata} \end{figure}

The MPFIT routine returns the measured shift, separately for $x$ and $y$, and
the error on these shifts determined from the $\chi^2$ minimization. Fig.
\ref{fig:rawdata} shows the reported 1-$\sigma$ uncertainties as function of the
factual measurements for both axes. The formal errors yield a typical S/N in
excess of 100 for our observed shifts. Significant clustering is visible in the
x-axis shifts, around the integer values. The measured values avoid the range
between ($-1.0$,$-1.1$) and ($-2.0$,$-2.2$). There may also be an excess at 0.
This appears to be a numerical artifact of the interpolation procedure, which is
part of the core IDL package. This artifact must be taken into account during
interpretation.

In the large majority of tiles, the uncertainty is less than 0.01 pixel, or 1
mas. The artifact discussed above amounts to errors of order 0.1 pixel, which
over a 10 year time span is still only 1 mas/yr. This is likely to be the best
achievable as the test was done under 'ideal' circumstances of no change in
emission apart from the motion. The real uncertainty is harder to quantify.
However, the spread in observed proper motions for different tiles, as shown in
Fig. \ref{fig:pmseparation}, provides an indication for this. The spread amounts
to an rms of 5 mas/yr, 5 times worse than derived from the numerical
uncertainty. The spread does not seem to increase with distance from the centre,
where the emission is fainter by two orders of magnitude. This indicates that we
are not limited by S/N on the emission, but by the presence of nebular
structures needed to detect proper motion.

\subsubsection{Results}
\begin{figure*} \centering \includegraphics[width=1. 
\textwidth]{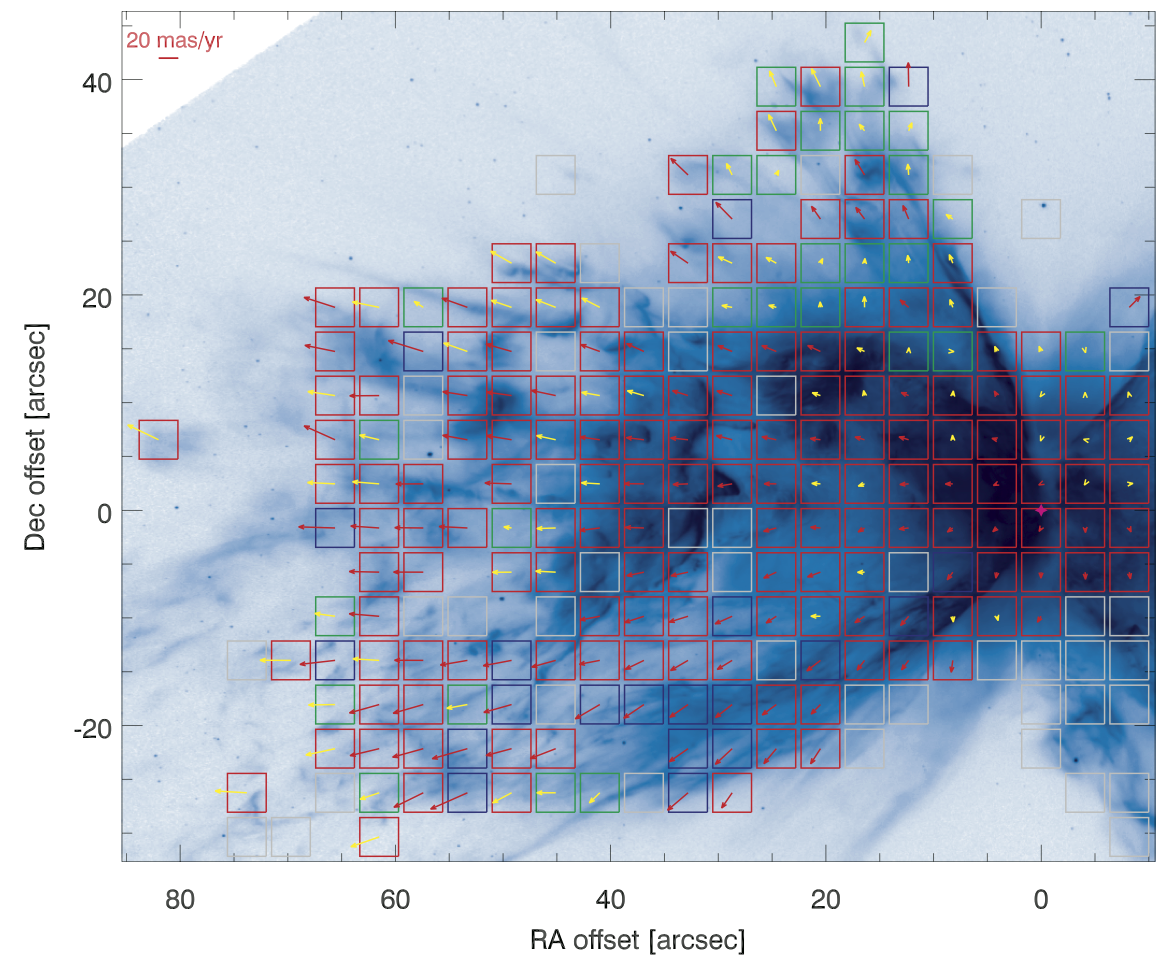} \caption{The proper motion velocity field
of NGC~6302, derived from the two HST images 9.44 yr apart. The blue image is
the 2009 WFC [N{\sc ii}] image.  Arrows indicate the measured proper motion for
each $4.1'' \times 4.1''$ tile. Yellow arrows indicate measurements where one of
the axes is in doubt. Red arrows give the most accurate results. The length of
each arrow indicates the proper motion; the bar at top left corresponds to a
value of 20 mas/yr. The typical uncertainty is estimated at 5 mas/yr. For empty
boxes, the measurement was classified as 'bad' and dropped. Colouring of boxes
corresponds to quality of the fit to the Hubble flow relation (see Fig.
\ref{fig:pmseparation}), with red points falling within 1-$\sigma $, blue at
least 1-$\sigma$ above the fitted relation and green at least 1-$\sigma$ below
the PM-separation relation. The reference position (RA offset, Dec.
offset)=(0,0) is the position of the central star RA=$17^h13^m44.39^s$ ,
Dec=$-37^{\circ}06'12.''93$ (J=2000) \citep{Szyszka:2009}.} \label{fig:vectors}
\end{figure*}

Fig. \ref{fig:vectors} shows the proper motion velocity field. The boxes show
the squares which were classified as 'good' by the automatic fitting routine on
the test shifts.  The arrows show the measured proper motion.  If the box is
empty, that specific measurement was flagged as 'bad' manually, either because
of the presence of a field star or because of lack of nebular structure.  The
yellow arrows indicate measurements where one of the axes returned a zero error;
this may in cases be related to the numerical artifact listed above. The
confidence in the direction of these vectors is reduced.  All remaining vectors
have good confidence. The length of each arrow is proportional to the proper
motion. The bar at the bottom right shows a proper motion of 20 mas/yr (0.2
pixels over 10 years). The magenta symbol at the centre of the nebula shows the
position of the central star from \cite{Szyszka:2009}.

A very regular velocity field emerges, largely radial from the location of the
central star \citep{Szyszka:2009}, at all distances from the star. The vectors
are well aligned with the nebular structures, in particular the bright linear
structures delineating the obscuring torus where the flow is along the
structures. The shock-like structure embedded in the centre of the eastern lobe
(from tile (37'', 15'') to tile (45'', -14'')) moves radially, at an angle to
the front, however where such a structure develops a tail (e.g. tile (55'',
18''), the tail is along the flow direction. Such tails are seen especially in
the outer regions, and appear to be a flow-driven phenomenon.

\section{Discussion}

\subsection{Hubble flow}
\begin{figure} \centering \includegraphics[width=0.45
\textwidth]{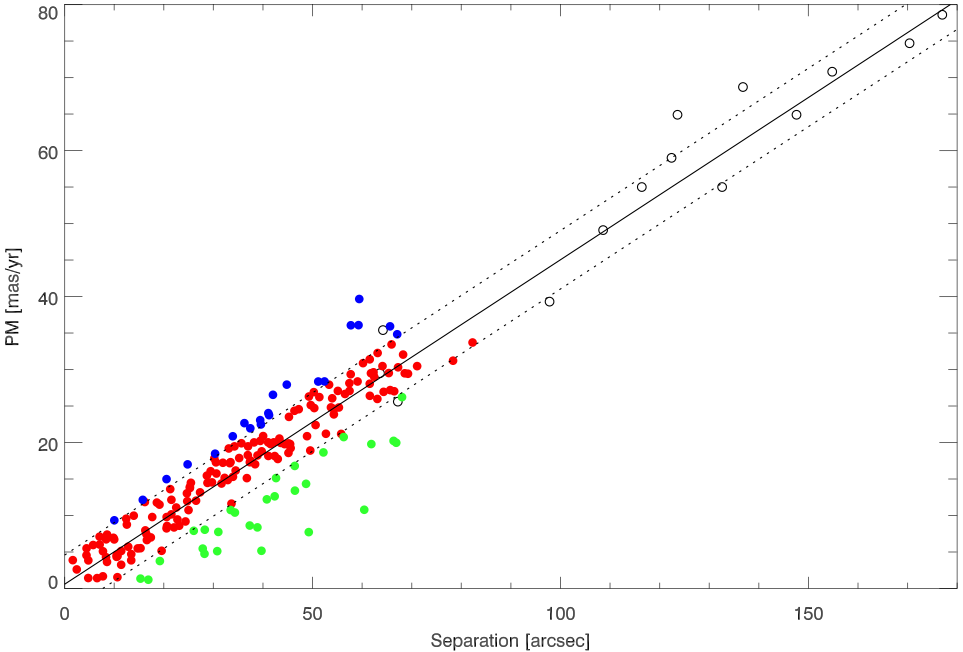} \includegraphics[width=0.45
\textwidth]{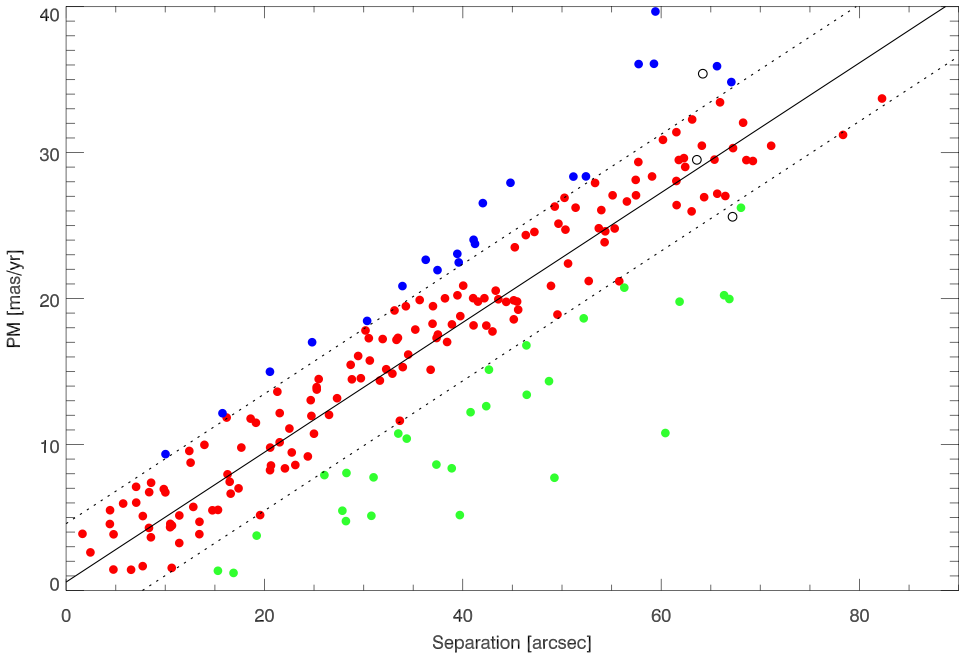} \caption{The proper motions derived for 200
individual tiles are plotted versus separation from the central source. Filled
circles represent this work; open circles from \citet{Meaburn:2008}. Note that
Meaburn's measurements are from the northwestern lobe of the nebula, while this
work concentrates mainly on the east lobe. The diagonal line represent the best
fit to all data (see Eq. (\ref{eq:1})) with an indication of $\pm$1-$\sigma$
dotted lines. Colouring of the points distinguish measurements falling within
$\pm$1-$\sigma$ form the fitted relation, form blue and red points which are
located at least  least 1-$\sigma$ above or below the fitted relation. This
colouring convention will be used in all subsequent figures.}
\label{fig:pmseparation} \end{figure}

So called Hubble-like outflows (viz. recession velocity $\propto$ distance) have
been found in a number of bipolar planetary nebulae
\citep{Corradi:2004,Jones:2010} and appear to be relatively common in such
objects.  In the case of NGC~6302, such a velocity law was first reported by
\citet{Peretto:2007}, as a component in molecular CO lines close to the
expanding torus: their Fig. 10 shows a clear linear dependence of velocity with
distance. A Hubble flow in the lobes was reported by \cite{Meaburn:2008}, based
on a tight correlation between proper motion of 15 individual knots and distance
to the central star. In 'normal' (non-bipolar) nebulae, the expansion velocity
increases outward because of the overpressure of the ionized region
\citep[e.g.][]{Gesicki:2003}. However, the linearly increasing Hubble flows are
more likely related to short-lived ejection events \citep{Huggins:2007} or fast
changes in outflow velocity \citep{Zijlstra:2001}.

In Fig. \ref{fig:pmseparation}, the dependence of the proper motion magnitude
versus separation of the individual knot from the central source is plotted for
our HST data (including all measurements presented in Fig. \ref{fig:vectors}). 
We assume that the central star is the point source revealed by the imaging in
the WFC3 F683N filter, at the position $\alpha = 17^h13^m44.39^s,\delta =
-37^{\circ}06'12.''93$ \citep{Szyszka:2009}. The data points from
\cite{Meaburn:2008} are presented in this figure by open circles. We find that a
linear relation provides an accurate representation of the results. The
measurements of Meaburn et al. fall on the same line, in spite of being located
on the oposite side of the nebula, and in most cases at much larger distances
from the star. The HST points extend to 70'' from the star, while the Meaburn
knots range from 60'' to 180''. Thus, both lobes have a common origin.

A linear fit is shown in Fig. \ref{fig:pmseparation}. The parameters of this
best fit, obtained from combining the HST and Meaburn et al. data (214 data
points), is given by \begin{equation}\label{eq:1} y = ax +b = 0.445 x +
0.58,(\sigma_{a}=0.01,\sigma_{b}=0.51) \end{equation} \noindent where $y$ is the
proper motion in mas/yr, and $x$ is the distance to the centre in arcsec. We
quantified the spread of points about the fitted line by calculation of the
standard deviation, which was $\sigma = 4.04$. This quantity is indicated in the
Fig. \ref{fig:pmseparation} by parallel dotted lines.

Fig. \ref{fig:pmseparation} shows a number of points falling somewhat below the
fitted line.  Fig. \ref{fig:vectors} shows the location of these vectors,
encoded by the green boxes.  Most of these vectors come from tiles with
relatively smooth emission, or from tiles only containing linear features along
the flow direction; all are low confidence points marked by yellow arrows. We
conclude that 'green' points can be safely omitted from the fit.  Refitting the
line excluding these lower points (186 points) yields:
\begin{equation}\label{eq:2} y = ax +b = 0.446 x +
1.67,(\sigma_{a}=0.007,\sigma_{b}=0.32) \end{equation} \noindent The blue
vectors show the points more than 1-$\sigma$ above the fit. These are
predominantly found in the lower right corner. Excluding also the higher points
we retain 165 data points which result in fit: \begin{equation}\label{eq:3} y =
ax +b = 0.444 x + 1.12,(\sigma_{a}=0.006,\sigma_{b}=0.31). \end{equation} The
fit remains the same within the errors, but the spread is significantly reduced.
Without the lower points, the spread is 2.9 mas/yr, which is a reasonable value
for the measurement error on good squares.

As the HST and Meaburn data are well fitted by a single line, we conclude that
the Hubble flow detected by \cite{Meaburn:2008} extends to the innermost parts
of the nebula.

\cite{Meaburn:2005} find that the outer lobe is tilted with respect to the plane
of the sky at an angle of 12.8$^{\rm o}$ (77$^{\rm o}$\ to the line of sight).
The current data does not restrict the angle of inclination for the inner lobe,
as the measured proper motion versus distance is inclination independent.  In
published archival spectra \citep{Meaburn:1980b,Meaburn:2005}, an additional
radial-velocity component is clearly visible in the inner regions attesting to
the presence of multiple components that may not occupy the same range of
inclination angles.  Thus, it is likely that a broader range of inclination
angles exists in the inner flow.

The Hubble like flow reported by \citet{Peretto:2007} in molecular CO lines is
detected in radial velocity, within the inner 20 arcsec. The increase of the
radial velocity is $\approx$\,24\,km/s over 15\,arcsec, or 280
km\,s$^{-1}$\,pc$^{-1}$, in the blue-shifted component. The Hubble flow reported
in this work gives a tangential velocity gradient of 410
km\,s$^{-1}$\,pc$^{-1}$. If these components trace the same outflow, the
inclination to the line of sight is 56$^{\rm o}$. The redshifted CO component
shows a minor Hubble flow with a similar gradient, indicating a similar
inclination to the line of sight. This symmetry is consistent with
\cite{Meaburn:2005} finding that the lobes are oriented close to the plane of
sky.

The velocity gradient of the Hubble flow derived from the tangential velocity,
410 km\,s$^{-1}$\,pc$^{-1}$, gives the correct value, as both velocity and
distance are projected by the same factor.

\subsection{Age of the nebula}
\begin{figure} \centering \includegraphics[width=0.45
\textwidth]{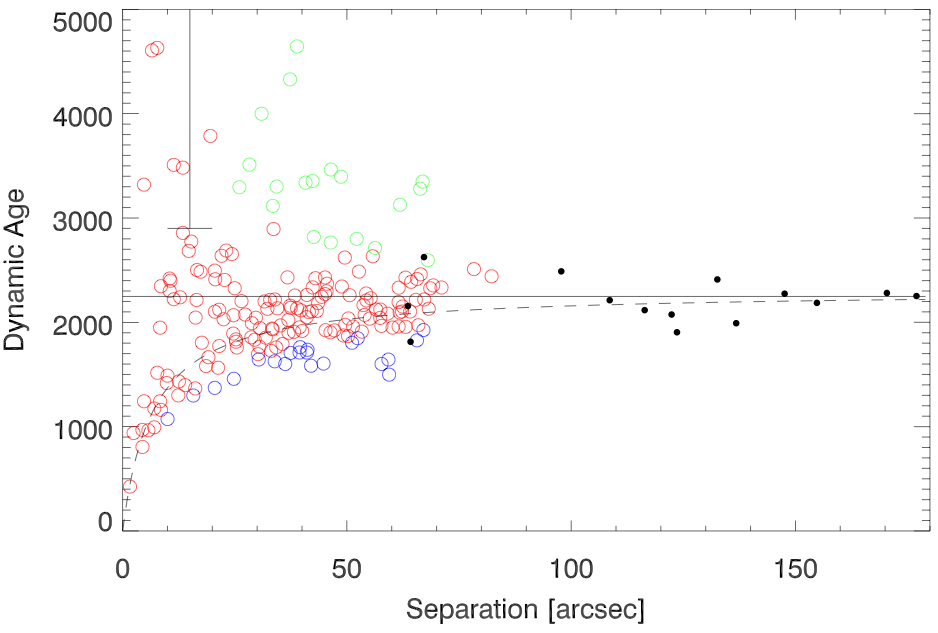} \caption{The ages of individual knots are plotted as
a function of distance from the central star. Models discussed in text
(Eq.~(\ref{eq:5})) are presented with lines. The solid line uses parameters from
the fit to all measurements (see Eq.~(\ref{eq:1})), while the dashed line
represent a fit without excluded low confidence points (see Eq.~(\ref{eq:2})).
The vertical bar indicates the mass-loss phase with cessation at 2900 yr
\citep{Peretto:2007}. The color of the points is as in Fig.~\ref{fig:vectors}. }
\label{fig:age} \end{figure}

\citet{Meaburn:2008} reported an age of $\approx$~2200 yr.  The global age of
the nebula can be derived from the slope of the Hubble flow relation. In this
work it varies between 0.44 and 0.45 mas\,yr$^{-1}$\,arcsec$^{-1}$ (see Eq.
(\ref{eq:1})-(\ref{eq:3})), which translates to an age of 2250 $\pm$ 35 yr.

The large number of individual measurements allows us to study the age of the
nebula in more detail.  The separation of specific fragments of the nebula from
the central star divided by their measured proper motion gives the dynamic age
of this patch of the nebula.  The distribution of these ages as function of
distance from the star is plotted in Fig. \ref{fig:age}.  If all the nebular
structures were ejected at the same time, and the velocity field remained
unchanged since the ejection, the distribution in Figure \ref{fig:age} would be
flat. This is not the case for NGC~6302. The scatter increases towards the
central star, because of the large effect of measurement errors. However,
allowing for this, there is a clear indication that the inner parts of the
nebula appear dynamically younger than the outer parts.

To explain this phenomena, let us make the assumption that the nebula
experienced an additional act of acceleration at some point during its
evolution. If this occurred very recently (compared to the time of ejection of
the AGB envelope), it will alter the nebular velocity field but not yet the
position of each specific fragment of the nebula. This will reduce the apparent
ages; the reduction will be largest in the inner part of the nebula where the
pre-acceleration velocities were the smallest.

We model this by adding an extra velocity component \begin{equation}\label{eq:4}
v= v_{0} + v_{c}. \end{equation} \noindent The modified age $t'$~[yr] can be
described by the formula \begin{equation}\label{eq:5} t' = \frac{10^{3}r}{v_{0}
+ v_{c}}= \frac{t_{0}}{1+{v_{c}t_{0}}/{10^{3}r}}, \end{equation} \noindent where
$r$~[arcsec] is the separation from the central star, $v_{0}$~[mas/yr] the
initial velocity and $v_{c}$~[mas/yr] represents the additional velocity
component. The conversion factor $10^{3}$ [mas/arcsec] is needed for
conservation of units. For small $v_{c}$, the modified age $t'$ becomes $t_{0}$
and is a constant function of separation. Eq.~(\ref{eq:5}) can be also presented
as a function \begin{equation}\label{eq:6} v(r) = \frac{10^{3}}{t_{0}}r + v_{c}
, \end{equation} \noindent which has the same form as the linear fit to the PM
data in Fig \ref{fig:pmseparation}. Thus, the fitted parameters in Eq.
(\ref{eq:1})-(\ref{eq:3}) represent $a = {10^3}/{t_{0}[\rm {1}/{yr}]}$ and $b =
v_{c}[\rm mas/yr]$.

The solid line in Fig. \ref{fig:age} shows this model for the fitted parameters
of Eq. (\ref{eq:1}). Excluding the low-confidence points above the relation (the
green points), as fitted in Eq. (\ref{eq:2}), gives the dashed line. The dashed
line gives a derived age $t_{0}$ of the nebula slightly less than before, of
2240 yr.

The best fit reveals an acceleration that occurred in the planetary nebula phase
and resulted in a velocity increase of $v_{c} = 9.2\,\rm km\,s^{-1}$. A similar
effect was observed in \citet{Gesicki:2003} where extra acceleration was needed
at the inner edge of a nebula to explain the velocity broadening of profiles of
high excitation ions.

The same kind of internal acceleration was reported by \citet{Peretto:2007} in
their figure 10. The Hubble flow derived from radial velocity CO observations at
the offset 0 [arcsec] does not have a zero velocity. Instead the slope of Hubble
flow relation starts at a velocity of $V = V_{\rm LSR} +12$\, km/s.

\subsection{Nebula asymmetries}
\begin{figure} \centering \includegraphics[width=0.45
\textwidth]{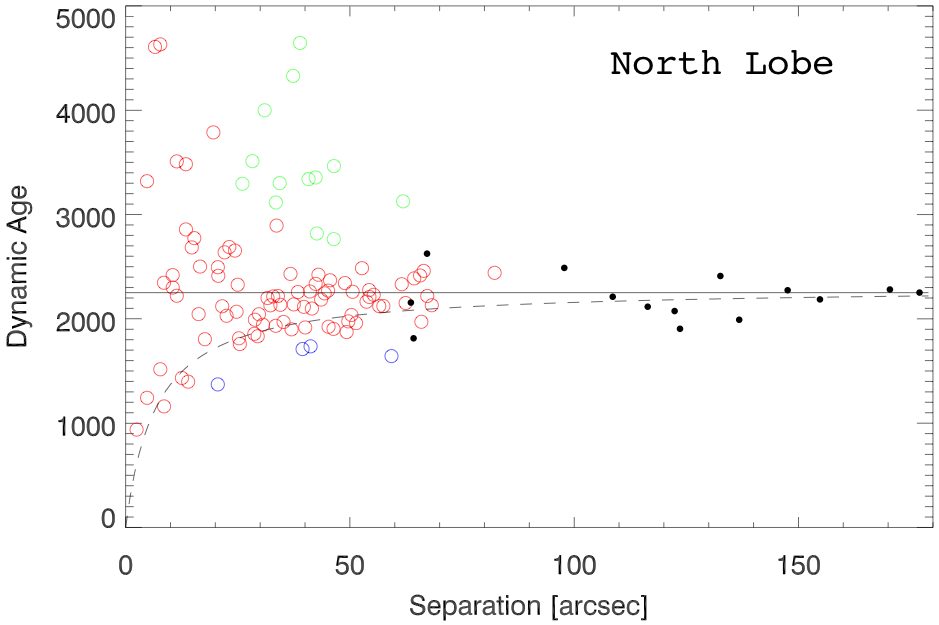} \includegraphics[width=0.45
\textwidth]{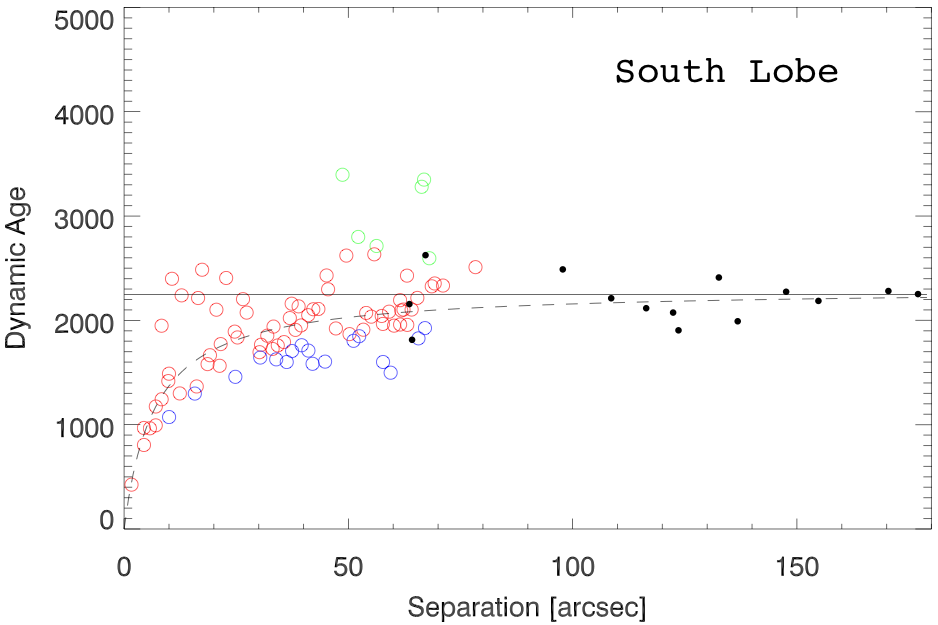} \caption{The same as Fig. \ref{fig:age}, but
northern tiles (DEC offset $\geq$ 0 in Fig. \ref{fig:vectors}) are plotted in
upper pannel and southern tiles of the nebula are plotted in lower panel. The
apparent age distribution differs in the northern and southern parts of the
nebula. The effect is attributed to additional acceleration present in the
southern lobe.  The dashed line represents the modified age $t'$ with parameters
from Eq. (\ref{eq:7}).  The fit to the southern tiles was made in PM-separation
space and is presented in Fig. \ref{fig:pmsouth}. The solid horizontal line
indicate an age of 2250 yr. The coloring of the points as in Fig.
\ref{fig:pmseparation}.} \label{fig:agesouth} \end{figure}

\begin{figure} \centering \includegraphics[width=0.45
\textwidth]{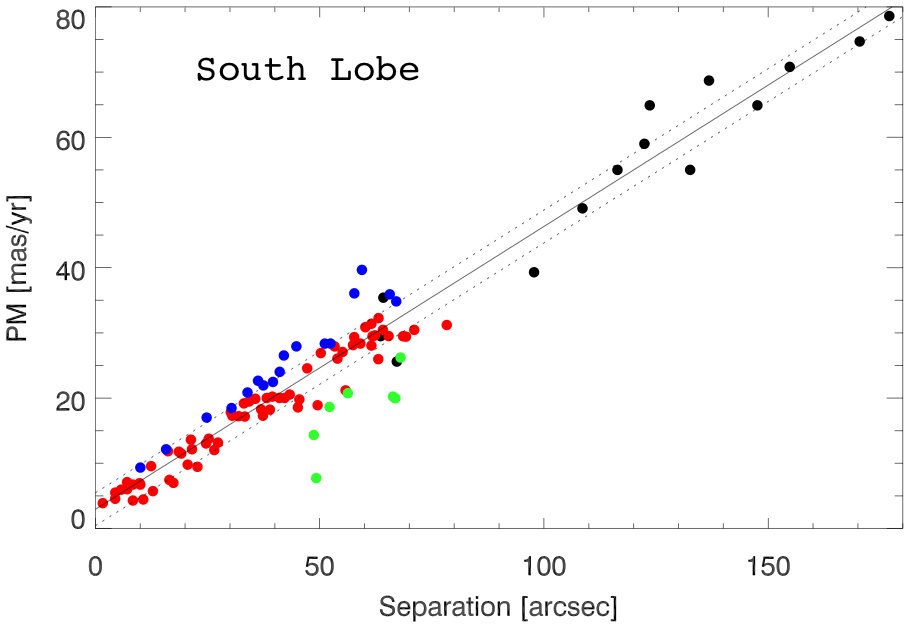} \caption{The same as Fig.~\ref{fig:pmseparation},
but only southern tiles (DEC offset \textless~0 in Fig.~\ref{fig:vectors}) are
plotted. The fit (solid line) is made to those measurements excluding outlaying
points (see Eq.~\ref{eq:7}), 1-$\sigma$ uncertainties are indicated with dotted
lines above and below. The coloring of the points as in Fig.
\ref{fig:pmseparation}.} \label{fig:pmsouth} \end{figure}

The consistency of the Hubble flow throughout the lobes indicates that the
expansion is homologous, and effectively free-flow. As derived in subsection
4.2, there is some evidence for additional acceleration; although this affects
the inner regions most, it may in fact be present throughout the nebula.

However, there appear to be some differences in kinematics between different
regions. The tiles where the gas is moving faster than expected (the blue points
in Fig. \ref{fig:pmseparation}) are primarily located in the southern part of
the nebula (the blue-edged tiles in Fig. \ref{fig:vectors}).

To show this, we divided the tiles into two samples, depending on the offset in
Fig. \ref{fig:vectors}. The northern tiles are defined as those where the
declination offset $\geq 0 "$ relative to the position of the central star and
the southern tiles are those with offset \textless~0''. The apparent age ($t'$)
distribution for both samples is presented in Fig. \ref{fig:agesouth}.  Very few
tiles from the southern part of the nebula appear to be older than 2250~yr,
compared to the northern tiles. The effect of the additional velocity component
is also much clearer in the southern part of the lobe. The differences are seen
mainly within 50 arcsec of the central star.

We fitted the Hubble-flow relation to the southern sub-sample of the nebula. We
excluded points falling outside the 1-$\sigma=4.04$ [mas/yr] upper and lower
threshold of the distribution, n the same way it was done for
Eq.~(\ref{eq:1})-(\ref{eq:3}). The re-fitted relation is

 \begin{equation}\label{eq:7} y = ax + b = 0.429 x +
 2.71,(\sigma_{a}=0.011,\sigma_{b}=0.7). \end{equation} \noindent The fit is
 presented in Fig.~\ref{fig:pmsouth}.  The parameter $b=v_{c}=2.71$\,[mas/yr],
 expressed as the velocity in the plane of the sky equals $v_{c} = 15\,\rm
 km/s$. This is a bit higher than previously found for the whole nebula.

The nature of these velocity asymmetries is not well understood. In
Fig.~\ref{fig:vectors} we notice a cluster of short vectors predominantly along
the northern edge of the eastern lobe. These represent the measurements with
large dynamical ages ($\geqslant$ 2700yr). In the southern section of the nebula
we observe higher velocities, thus younger dynamical ages, at the same distance
from the star. This cannot be due to a difference in inclination with the line
of sight, as this projection effect disappears in the Hubble diagram. One
possibility is that the northern region has undergone deceleration; another
possibility is that the innermost southern region is younger.  A plausible model
for the development of multipolar nebulae is the warped disk model of
\citet{Icke:2003}, suggested to be applicable to NGC\,6302 by
\citet{Matsuura:2005}.  In this model, the most elongated lobes develop only
once the stellar wind breaks through the constraining disk. This happens after a
significant time delay. It is noteworthy that the southern, younger vectors in
NGC\,6302 are found point-symmetrically to the most elongated north-western lobe
\cite{Meaburn:2005}.

\subsection{Centroid of PM measurements } With a sufficient number of PM vectors
we can attempt to find the point of common origin of these velocities. Two
methods were used to test whether the PM vectors converge in one point. The
first is a propagation of each measurement back in time, conserving its
direction. The time is derived from the distance to the assumed central
position. Each vector gives a position at the same distance as assumed for the
central star position. The second is a more novel technique called {\it
criss-cross mapping} \citep{Steffen:2010b}, and relies solely on directions of
vectors.

\begin{figure} \centering \includegraphics[width=0.45
\textwidth]{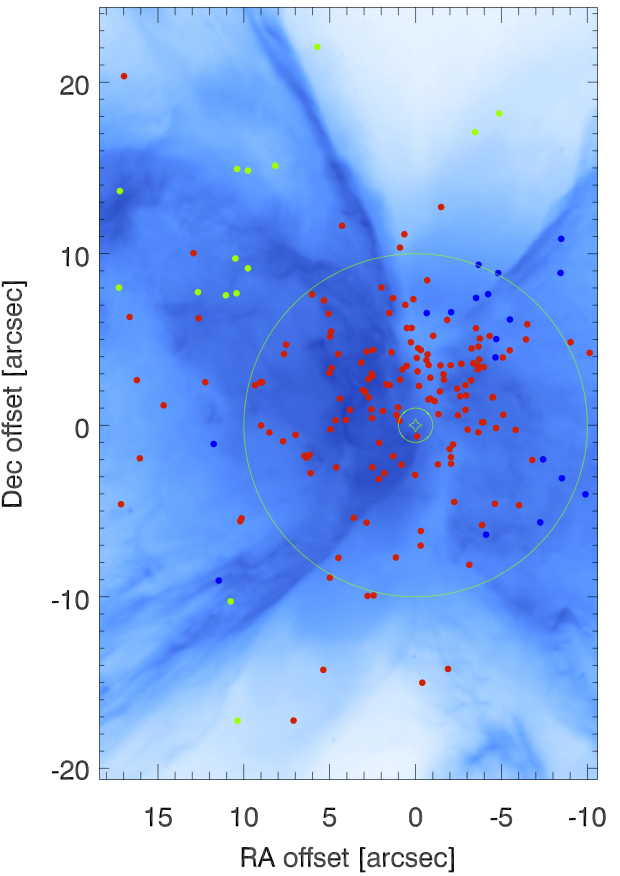} \caption{The origins of the PM vectors are plotted
against inner parts of the NGC~6302 (blue image). For each fragment of the
nebula at the distance $r$ we adopted age $t'$ according to Eq. \ref{eq:5}, and
fitted values from equation (\ref{eq:2}) $v_{c}=1.67$~[mas/yr] and age $t_{0} =
2220$~[yr]. The position of the central star is indicated by the green star. The
1'' and 10'' circles are plotted for the reference. The (0,0) point corresponds
to the position of the central star. North is up, East is left.}
\label{fig:origins} \end{figure}

In the first approach we propagated the PM vector in the opposite direction to
the measured motion. We assumed that the age $t'$ [yr] for each specific
fragment of the nebula at the distance $r$ [arcsec] can be described by Eq.
(\ref{eq:5}) using parameters from the fit to the Hubble flow relation (see Eq.
(\ref{eq:2})). The derived age together with the PM vector defines the 'origin'
point of this specific part of the nebula.  The result is presented in Fig.
\ref{fig:origins}, with the same colouring conversion as in Fig.
\ref{fig:pmseparation}. The background is the optical image of the nebula.  We
can see that the convergence is rather poor.  Out of 200 considered PM
measurements, 155 fall within the circle of 10 arcsec from the central star.
These are mostly (133) 'red' points which follow the Hubble-flow relation
reasonably well (see Fig. \ref{fig:pmseparation}). We also find 11 'blue' origin
points above this relation; notably, none of the 'green' points was found within
the 10'' circle.

\begin{figure} \centering \includegraphics[width=0.45
\textwidth]{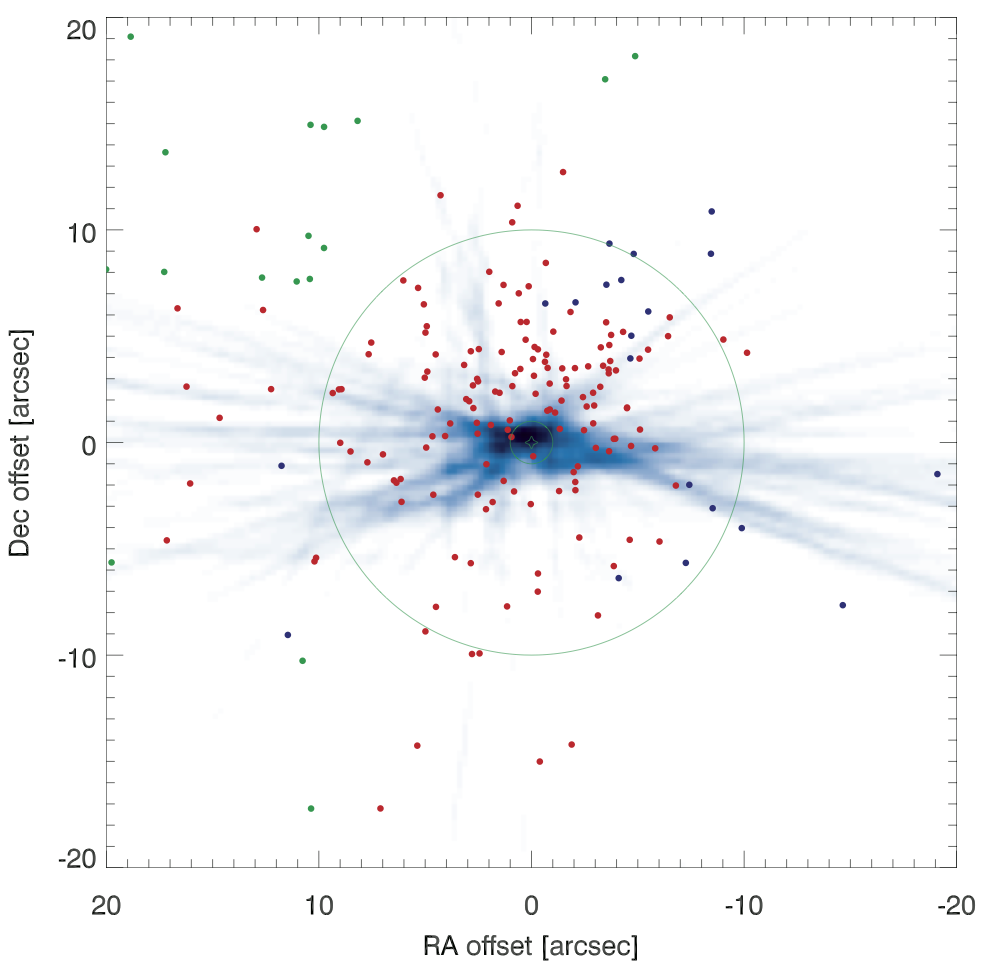} \caption{The criss-cross mapping of the PM
vectors (the blue image) compared to the origins of the PM vectors calculated as
discussed in text and Fig. \ref{fig:origins}. The blue background is the
criss-cross map constructed from measured PM vectors.  The colouring of the
points is as in Fig. \ref{fig:pmseparation}. The and 10'' circle is plotted for
the reference. The (0,0) point corresponds to the position of the central star. 
North up East left. } \label{fig:crisscross} \end{figure}

The criss-cross mapping is implemented in the modelling tool Shape
\citep{Steffen:2010a}. In this technique we construct a map of crossing points
between all the measurements, much like an intensity map.  Each PM vector is
represented as an infinite line, with some width. At the crossing point the
number of lines is added up.  The peak of the the number density map indicates
the position where the largest number of PM vectors crossed each other.

The result of criss-cross calculation is presented in Fig. \ref{fig:crisscross}.
The number density map is squared to highlight regions with the highest
probability to be a centroid of the PM velocity field. The origin points from
the previous method are over-plotted for reference.  The convergence is very
close to the central star position (the (0,0) offset on the map), with an offset
of only $\approx$~0.5'' north. The criss-cross analysis provides support for the
identification of the point source found by \citet{Szyszka:2009} as a central
star. The fact that most of PM vectors cross the position of the central star
indicates that the velocity field is mostly radial.

The two methods show different results. The fact that for the first method, the
PM vectors do not converge to a common origin point, we attribute to the
variations in velocities due to the kinematic effects discussed above.  The
second method showed that a large number of the PM vectors indeed cross at the
position of the central star.  If we apply this result in the first method, we
conclude that the age was underestimated (for the northern lobe) or
overestimated (for the southern lobe).

 \subsection{The molecular torus and ionized centre}

The nebula has a prominent high extinction torus \citep{Matsuura:2005}. This
torus is also detected in CO, is oriented N-S, extends from 5.5'' to 12''
\citep{Peretto:2007}, and has an estimated dust mass of $M_{\rm dust} \sim
0.03$~M$_{\odot}$ and a total mass of $M_{\rm gas} \approx$~1~M$_{\odot}$
\citep{Matsuura:2005}. (\citet{Peretto:2007} improved the torus mass estimation
to $\approx$~2~M$_{\odot}\pm1$~M$_{\odot}$.  \citet{Dinh-V-Trung:2008} however
derive a much lower mass of 0.1\,M$_\odot$, based on almost identical data. This
value is surprisingly low given the extremely high extinction
\citep{Matsuura:2005}, but it shows that the mass determinations remain
uncertain.)

The molecular torus reported by \citet{Peretto:2007} is centered at declination
$\delta = -37^{\circ}06'12.''5$. The declination of the central star is $\delta
= -37^{\circ}06'12.''93$, whilst the peak of the criss-cross map is about 0.5''
North at $\delta = -37^{\circ}06'12.''5$.  All these positions agree to within
the uncertainties. The torus has an age of 2900\,yr, based on its size and the
expansion velocity of 8\,km\,s$^{-1}$, \citep{Peretto:2007}.

The region inside of the molecular torus is not devoid of gas.  There is an
inner, ionized torus, located within the central few
arcseconds\citep{Gomez:1989}. This ionized torus is also expanding
\citep{Gomez:1993}, at a fractional growth rate of
8.36$\,\times10^{-4}\,$yr$^{-1}$, which gives a dynamic age of 1200yr, making it
much younger than any other part of the nebula.  There are uncertainties in this
measurement. The ionization front travels faster than the gas by as much as
40\%\ and this can lead to overestimation of measured expansion rates of
ionization-bounded nebulae, as found for the case of NGC~7027 by
\citet{Zijlstra:2008}.  (\citet{Gomez:1993} apply their expansion to the inner
radius, assuming a Hubble-type velocity field but this retains the uncertainty.)
A second problem is that the velocity field in the ionized region can be altered
by the overpressure caused by the ionization, leading to acceleration. The
nature and evolution of the inner torus is not yet fully understood. The
expansion of the inner torus appears to be in the same direction as the faster
(younger) proper motion vectors measured by HST in the southeastern part of the
nebula, and there may be a relation between these two regions.

The HST data do not allow us to study the expansion of the inner 4'' of the
nebula as this is approximately the size of the single tile which is
4.1$''\times$4.1$''$. Because of the lack of clear features in this highly
obscured region, we have also been unable to make an expansion map of the inner
regions at higher spatial resolution.

\subsection{ Mass-loss history of NGC~6302}  We find three separate
components in NGC~6302: the lobes, expanding fast with a Hubble flow, the
molecular torus, expanding slowly at $\sim 8\,\rm km/s $, and the inner ionized
torus. This raises the question whether these components date from separate
mass-loss events, or have a common origin.

\citet{Peretto:2007} derived the timing and duration of the mass loss event
which lead to the formation of the torus. This event lasted $\approx$4600 yr,
began 7500 years ago and finished $\approx$2900 years ago. Fig. \ref{fig:age}
indicates the duration of this mass loss phase as a vertical bar. The present
work estimates the age of the lobes to be $\approx$2240 yr.  Thus, the lobes
formed after the torus was ejected, and there appears to be a delay of $\sim
650$\,yr between the ejection of the torus and lobes. The inner torus has a
reported age of 1200\,yr, but this needs confirmation.

Differences in time between the ejection of a torus and of a jet are known from
other objects. \citet{Huggins:2007}, based on a sample of 9 nebulae for which
both times where available, found that there was a small delay of typically a
few hundred years between torus and jets. For this time delay pattern he coined
the term {\it jet lag}. The fact that this statistical effect is confirmed in
this detailed study of the lobes and torus of NGC\,6302 provides support for
this mass-loss description.

Excluding the innermost torus, we find a mass-loss history for NGC\,6302
consisting of a slow, dense, equatorial wind with mass-loss rates of $\dot M
\sim 5 \times10^{-4}\,\rm M_\odot\, yr^{-1}$, lasting for roughly 5000\,yr,
followed by a brief interlude, before a short-lived event caused the formation
of the fast, bipolar lobes.

Assuming a lobe mass of $\sim 0.5\,\rm M_\odot$ \citep{Dinh-V-Trung:2008}, the
mass-loss rate of the rapid lobe ejection would seem to be extreme. However, it
is more likely that this event accelerated gas previously ejected.

\citet{Peretto:2007} argue that the momentum of the torus is almost an order of
magnitude larger than that in the stellar radiation field. The lobes, moving
much faster, having a few times less mass but having formed at least ten times
faster, have a much higher ratio of momentum to that in the stellar radiation
field over the time of the ejection. This suggests the late ejection was
energy-driven, as has been suggested for other bipolar post-AGB stars
\citep{Bujarabal:2001}.

\citet{Bujarabal:2001} suggest as one possibility that the fast outflows are
driven by conversion of gravitational energy of re-accreted material into
bipolar jet momentum. This model can also be proposed for NGC\,6302.
\citet{Peretto:2007} suggest that the torus ejection was aided by a binary
companion. In this case, the later fast flows could be powered by an accretion
disk around the same binary companion.

\section{Conclusions} The comparison of two HST F658N images allowed for
measurement of the PM vector for 200 4.1''$\times$4.1'' square tiles
predominantly across the eastern lobe.  The resultant velocity field is very
regular and largely radial.  We have shown that the Hubble-flow relation extends
into the innermost parts of the nebula. The age of 2250 yr of the nebula agrees
with the one derived for northwestern lobe by \citet{Meaburn:2008}.  The inner
parts of the nebula show evidence of additional acceleration which took place
only in the recent history of the nebula. The additional velocity component is
more clearly pronounced in the southern half of the lobe, than in the northern. 
We attribute this recent acceleration to overpressure after the onset of
ionization.

The criss-cross mapping indicates that the PM velocity field points toward the
central star reported by \citet{Szyszka:2009}. It also shows that the lobes and
the torus have the same kinematic origin, close to the central star.  The
comparison of torus ejection time with lobe ejection time provide further
evidence that these two events, although separated in time, are linked together.

The mass-loss history for NGC\,6302 shows a phase of equatorial, slow mass loss
at a high rate of $\dot M \sim 5 \times 5 \times 10^{-4}\,\rm M_\odot\,yr^{-1}$,
lasting about 5000\,yr, and followed by an interlude of $\sim 650\,$yr before
the fast lobes formed during a short-lived, energetic event. We suggest that the
lobes may have formed through an accretion disk around a companion to the
central star.

\section*{acknowledgements} We would like to thank W. Steffen for his assistance
preparing the criss-cross map. We also thank the anonymous referee whose careful
reading of the manuscript lead to improvements of this paper. CS is grateful to
the University of Manchester, School of Physics and Astronomy for a bursary.

This research made use of Montage, funded by the National Aeronautics and Space
Administration's Earth Science Technology Office, Computation Technologies
Project, under Cooperative Agreement Number NCC5-626 between NASA and the
California Institute of Technology. Montage is maintained by the NASA/IPAC
Infrared Science Archive.

\bibliographystyle{mn2e} \bibliography{resub_expansion_pm}

\end{document}